\documentclass[aps,prb,amssymb,12pt,floatfix]{revtex4}
\usepackage{subfigure}
\usepackage{graphicx}
\usepackage{rotating}
\usepackage{amsmath, amsthm}

\begin{document}

\title{Size, shape, and flexibility of RNA structures}
\author{Changbong Hyeon$^1$ and Ruxandra I. Dima$^2$ and D. Thirumalai$^{1,3}$}
\affiliation{\em \small $^1$Biophysics Program, Institute for Physical Science 
and Technology,
University of Maryland, College Park, MD 20742\\
$^2$Department of Chemistry, University of Cincinnati, Cincinnati OH, 45221\\
$^3$Department of Chemistry and Biochemistry, University of Maryland, College 
Park, MD 20742\\
}

\date{\small \today}
\baselineskip = 19pt
\begin{abstract}
Determination of sizes and flexibilities of RNA molecules is important
in understanding the nature of packing in folded structures and in
elucidating interactions between RNA and DNA or proteins.  Using the
coordinates of the structures of RNA in the Protein Data Bank we find that 
the size of the
folded RNA structures,  measured using the radius of gyration,
$R_G$,  follows the Flory scaling law, namely, $R_G =
5.5 N^{1/3}$ \AA\ where N is the number of nucleotides.  The shape of RNA molecules is characterized by the
asphericity $\Delta$ and the shape $S$ parameters that are
computed using the eigenvalues of the moment of inertia tensor.  From
the distribution of $\Delta$, we find that a large fraction of folded
RNA structures are aspherical and the distribution of $S$ values shows
that RNA molecules are prolate ($S>0$).  The flexibility of folded
structures is characterized by the persistence length $l_p$.  By
fitting the distance distribution function, $P(r)$ that is computed
using the coordinates of the folded RNA, to the worm-like chain model
we extracted the persistence length $l_p$.  We find that $l_p \approx
1.5 N^{0.33}$ \AA\ which might reflect the large separation
between the free energies that stabilize secondary and tertiary structures.  The dependence of
$l_p$ on $N$ implies the average length of helices should increases as the size of RNA grows.  
We also analyze packing in the
structures of ribosomes (30S, 50S, and 70S) in terms of $R_G$,
$\Delta$, $S$, and $l_p$.  The 70S and the 50S subunits are more
spherical compared to most RNA molecules.  The globularity in 50S is
due to the presence of an unusually large number (compared to 30S
subunit) of small helices that are stitched together by bulges and
loops.  Comparison of the shapes of the intact 70S ribosome and the
constituent particles suggests that folding of the individual
molecules might occur prior to assembly.
\end{abstract}
\maketitle

{\bf INTRODUCTION}\\

Molecular recognition between RNAs or RNA and protein is involved in a
number of cellular functions.  In all these processes RNA interacts
with other biomolecules.  In order to understand the biophysical basis of 
interactions of RNA with other biological molecules it is necessary to characterize the shapes of the
interacting partners.  Hence, it is important to elucidate the
shapes and flexibilities of RNA structures.  The large increase in the
number of three dimensional structures allows us to quantify RNA
shapes which is needed to describe the assembly of complexes such as the ribosome.

In contrast to the situation in RNA much is known about packing and
shape fluctuations in proteins.
\cite{RichardsQRB94,RichardsJMB74,FinneyJMB75,DillBJ01}  In part this
is because the number of solved protein structures is $\sim$30,000
while the RNA structure database contains only $\sim$600 structures.
Despite considerable success in the secondary structure predictions of
nucleic acid sequences using energy minimization dynamic programming
algorithm \cite{Zukerbook,HofackerNAR03} or comparative sequence
analysis \cite{GutellCOSB02} the complicated nature of
counterion-mediated tertiary interactions in RNAs makes it difficult
to obtain three dimensional RNA structures using computational
methods. The recent experimental determination of medium to large size
of RNA structures has prompted us to perform a statistical analysis of
RNA structures with the aim of characterizing their shapes and
flexibility.

In this paper, we study the structural features of RNA using the
currently available RNA three dimensional structures. \cite{PDB}  The
size of RNA, as measured by the radius of gyration $R_G$, shows that
typically RNA molecules are compact.  The variation of $R_G$ with the
number ($N$) of nucleotides obeys Flory law i.e., $R_G=aN^{1/3}$ \AA.
Although the overall scaling law for $R_G$ for RNA is identical to
that for proteins there are considerable differences in their shapes.
We find that the folded states of RNAs are largely prolate and are
considerably more aspherical than proteins.  The flexibility of RNA,
which is crucial in describing interactions with proteins and RNA and
DNA, is described in terms of the persistence length ($l_p$) which can
be measured using X-ray scattering \cite{CaliskanPRL05} and other
methods. \cite{BustamanteSCI94}  The values of $l_p$ for RNA, which
are considerably larger than for proteins, vary between (5-30)\AA\
depending on $N$.  Using the shape parameters and $l_p$ we also
describe the unusual structural characteristics of the ribosome, a
large ribonucleoprotein complex. \\

{\bf METHODS}\\

{\bf RNA structures :} We computed several quantities to characterize
the shapes of RNA using the atomic coordinates of their structures
determined by X-ray crystallography, NMR, or cryo-EM.  The coordinates
for all RNA structures were obtained from the Protein Data Bank (PDB).\cite{PDB}  Our analysis is performed for over 1185 individual RNA
chains with the number of nucleotides $N>10$ found in 642 RNA related
PDB files as of June 2005.  Among these, 195 RNA chain structures are
monomers, and the rest of the chains are part of oligomers  or
appear in complexes with other RNA molecules or proteins.  Structural
features in the monomeric form can be different from those determined
in an oligomer or complex because the intermolecular interaction can
affect the individual chain structure.  Therefore, we analyzed the two
groups of structures separately.  For comparison, we have also
calculated shape characteristics for a dataset of proteins.  The
results for proteins enable us to assess certain unusual features of
RNA-protein interactions especially in the ribosome.\\

{\bf Size :} The radius of gyration ($R_G$) is an indicator of the
overall size of RNA.  The value of $R_G^2$, which can be measured
using small angle X-ray or neutron scattering, is calculated using
\begin{equation}
R_G^2=\frac{1}{2\sum^M_im_i\sum^M_jm_jN^2}\sum_i^M\sum_j^Mm_im_j(\vec{r}_i-\vec{r}_j)^2.
\label{eqn:RG}
\end{equation}
where $M$ is the number of atoms in the molecule, and $m_i$ is the mass of the $i^{th}$ atom. In the calculation of $R_G^2$ for RNA structure we used only the
coordinates of the heavy atoms (C, N, O, P). \\

{\bf Shape :} The deviation from the spherical shape is characterized
by the asphericity $\Delta$ and the shape parameter $S$, both of which
are calculated from the inertia tensor,\cite{NelsonJP86,HoneycuttJCP89}
\begin{eqnarray}
\mathcal{T}_{\alpha\beta}&=&\frac{1}{2\sum^M_im_i\sum^M_jm_j}\sum^M_i\sum^M_jm_i
m_j(r_{i\alpha}-r_{j\alpha})(r_{i\beta}-r_{j\beta})\nonumber\\
&=&\frac{1}{\sum_i^Mm_i}\sum_{i}^Mm_i(r_{i\alpha}-R_{C\alpha})(r_{i\beta}-R_{C\beta})
\end{eqnarray}
where ${r}_{i\alpha}$ is
the $\alpha$-th Cartesian component of the position of atom $i$, and
$\vec{R}_{C}=\sum_{i}^Mm_i\vec{r}_{i}/\sum_{i}^Mm_i$ is the a center
of mass.  The square of the radius of gyration is
$R_G^2=tr\mathcal{T}$.  The eigenvalues $\lambda_1$, $\lambda_2$ and
$\lambda_3$ of the matrix $\mathcal{T}$ are the the squares of the
three principal radii of gyration.  The extent of asphericity is
characterized using $\Delta$ ($0\leq \Delta \leq 1$)
\begin{equation}
\Delta = 
\frac{3}{2}\frac{\sum_{i=1}^3(\lambda_i-\overline{\lambda})^2}{(tr\mathcal{T})^2
}
\label{eqn:asphericity}
\end{equation}
where $\overline{\lambda}=(\lambda_1+\lambda_2+\lambda_3)/3$.
For a perfect sphere $\Delta=0$.
Deviation from $\Delta=0$ indicates the extent of anisotropy.
The overall shape of a molecule is assessed using
\begin{equation}
S = 27\frac{\prod_{i=1}^3(\lambda_i-\overline{\lambda})}{(tr\mathcal{T})^3}
\label{eqn:shape}
\end{equation}
which satisfies the bound $-1/4\leq S \leq 2$.  Negative values of $S$
correspond to oblate ellipsoids and $S>0$ are prolate ellipsoids.

Most studies of packing in proteins and RNAs involve tessellation of
space which always introduces certain amount of arbitrariness.\cite{DillBJ01,GersteinJMB05}  
In contrast, the shape parameters
$\Delta$ and $S$ are directly computed using only the atomic
coordinates.  Knowledges of $\Delta$ and $S$ are important in
determining the overall motion of RNA and their interaction with other
biomolecules.  \\

{\bf Persistence Length :} A parameter that describes the
flexibility of biomolecules is the persistence length, $l_p$, which is
most clearly defined by assuming that RNA structures can be described
by a polymer model.  Based on previous experimental studies it is
suspected that the statistical properties of dsDNA,\cite{BustamanteSCI92,BustamanteSCI94} ssDNA,\cite{WeillMACRO97} and
RNA \cite{Bustamante2,CaliskanPRL05} can be described using the
worm-like chain (WLC) model.  For WLC models $l_p$ can be estimated
provided the distribution of the mean end-to-end distance $R_E$ or
$R_G$ is known.  Exact calculation of neither $P(R_E)$ nor $P(R_G)$ is
possible for WLC.  A simple and accurate theoretical expression has
been derived for $P(R_E)$ of worm-like chain using the mean field
approximation.\cite{Habook,HyeonJCP06}  The resulting distribution,
which is in good agreement with computer simulations,\cite{FreyPRL96} is
\begin{equation}     
P_{WLC}(r_E) = \frac{4\pi Cr_E^2}{(1-r_E^2)^{9/2}}\exp[-\frac{3t}{4(1-r_E^2)}].
\label{eqn:WLC}
\end{equation}
where $r_E\equiv R_E/L$ and $t\equiv L/l_p$. $L$ is the contour
length.  For RNA molecules, which from the perspective of polymers,
can be viewed as a branched polyelectrolyte chains, the contour length
is also an unknown parameter.  The normalization constant
$C=1/[\pi^{3/2}e^{-\alpha}\alpha^{-3/2}(1+3\alpha^{-1}+15/4\alpha^{-2})]$
with an $\alpha=3t/4$.  When $l_p$ is small $P_{WLC}(r_E)$ reduces to a
Gaussian chain whereas for large $l_p$ $P_{WLC}(r_E)$ approaches the
rod-limit as $r_E\rightarrow 1$.

Although  direct measurements of $P(R_E)$ for  biomolecules are not routinely performed it is conceivable that $P(R_E)$ may be obtained using single
molecule FRET experiments.  However, the distance distribution
function $P(r)$ can be measured using SAXS experiments.\cite{SosnickBC00,RussellNSB00,HerschlagPNAS02}  Based on general
arguments, we expect that the distribution functions $P(r)$ and
$P(R_E)$ should coincide provided $r\gg R_G$. Because $\langle
R_E^2\rangle\sim\langle R_G^2\rangle\sim Ll_p$ for WLC provided $L$ is
large it follows scaling arguments that $P(r)$ should decay for large
$r$ as
\begin{equation}
P(r)=\beta\exp\left(-\frac{1}{1-x^2}\right)
\label{eqn:fit}
\end{equation}
where $x=l_pr/R_G^2$, and $\beta$ is an arbitrary constant. In
practice Eq.\ref{eqn:fit} accurately describes $P(r)$ computed using
the coordinates of RNA structures when $r/R_{g} > 1$.  We determined
$l_p$ by fitting the $P(r)$ function for RNA structures to
Eq.\ref{eqn:fit}.

Recently, we used Eq.\ref{eqn:fit} to analyze small angle X-ray
scattering data. We showed that $l_p$ for the Azoarcus ribozyme
changes by a factor of 2 as the molecule folds upon addition of
counterions (Mg$^{2+}$ or Na$^{+}$). Although the structural basis for
the success of WLC in describing certain properties of folded RNA is
unclear, Eq.\ref{eqn:fit} is useful in analyzing scattering data.

For purposes of comparisons we have also calculated $P(r)$ for folded
structures for 56,000 protein chains.  To our knowledge the
persistence length of proteins has not been directly measured. We
obtain $l_p$ by fitting $P(r)$, obtained from the coordinates of the
structures in the PDB, to Eq.\ref{eqn:fit}.\\

{\bf RESULTS}\\

{\bf Distribution of RNA structures as a function of N:} From the
distribution of $P(N)$ the number of RNA structures in the PDB as a
function of chain length ($N$) in Fig.\ref{PDB_RNA} we find that $\sim
70$\% of the database contains $N$ in the range $10<N<30$.  The peak
in $P(N)$ between $70<N<80$ is due to the large number of tRNA
structures that have been determined in various conditions.  The peaks
at $N\approx 1500$ and $N\approx 3000$ correspond to 16S and 23S
ribosomal RNAs, respectively.  Compared to statistics of protein
structures (see Fig.\ref{PDB_RNA} inset), RNA structures are more
clustered at small values of $N$ but span a broader range of $N$.
However, this distribution is unrelated to the number of RNA
molecules that are relevant to biological functions.  There is a broad
range in $N$ that represents noncoding RNAs.  For example, the length
of human ncRNA functioning in gene silencing process is $\sim$100,000
nucleotides.\cite{BarlowNature02}  From Fig.\ref{PDB_RNA}, which reflects the
current status in RNA structure determination, it is clear that there
is a large gap between the total number of functional RNAs and those
with known three dimensional structures. \\

{\bf Size of RNA obeys the Flory law :} If the overall shape of RNA is
spherical then its volume, an extensive variable, is
$V \approx \frac{4\pi}{3}R_G^3$ with $R_G$ being the radius of gyration. 
For accurate computation of volumes one should use the hydrodynamic radius instead of $R_G$.
Because $V\sim a^3N$ where $a$ is a characteristic length
(approximately the distance separating two consecutive nucleotides) it
follows that $R_G\sim aN^{1/3}$.  This general result was first
derived by Flory who showed that $R_G\sim aN^{\nu}$ where $\nu=1/3$
for maximally compact structures.  Because RNA is a polyelectrolyte
its $R_G$ depends on the concentration of counterions ($C$).  At low
values of $C$, RNA is expanded and the transition to a compact
structure occurs only when $C$ exceeds a critical value.

We calculated $R_G$, using Eq.\ref{eqn:RG} (see Methods), for the 1155
``folded'' RNA structures. A plot of $R_G$ as a function of $N$
confirms the Flory result.  From the plot in Fig.\ref{RNA_Rg_N} we
find that, for the folded RNA structures, $R_G$ can be accurately
calculated using
\begin{equation}
R_G=aN^{1/3}
\label{eqn:Rg_vs_N}
\end{equation}
where $a=5.5$\AA.  The prefactor, $a=5.5$\AA, for the folded
structures approximately corresponds to the average distance
($\approx$5.5\AA) between the phosphate groups along the backbone
(Fig.\ref{P-P}).  Recent measurement of $R_G$ for the compact state of
the 195 nucleotides \emph{Azoarcus} ribozyme at high concentration of
$\mathrm{Na^+}$ or $\mathrm{Mg}^{2+}$ shows that $R_G\approx 35$\AA.
\cite{CaliskanPRL05}  From Eq.\ref{eqn:Rg_vs_N} we find $R_G\approx
32$\AA.  This analysis further suggests that the prefactor in
Eq.\ref{eqn:Rg_vs_N} may indeed be interpreted as the mean distance
between consecutive phosphate groups in the folded structures.  If the $R_G$ data in Fig.\ref{RNA_Rg_N} 
for $N<20$ is neglected we find that Eq.\ref{eqn:Rg_vs_N} is obeyed with $a \approx 5$\AA. Thus, the scaling relation is robust.

It is perhaps more reasonable to view RNA structures as formed from
relatively rigid duplexes that are linked by flexible motifs such as
bulges, loops, etc. In such a picture the fraction of base-paired
nucleotides can be chosen as a variable to describe the overall
size. We have shown previously (see Fig. 10 in \cite{DimaJMB05}) that
the number of base pairs in RNA is $\propto$ N. Thus, the Flory result
would be valid even if one accounts for the rigidity of RNA duplexes. \\

{\bf Single-chain RNAs are aspherical and prolate :} Even though
folded RNA structures are compact, as assessed by their size, there
are substantial deviations from sphericity.  Indeed, the distribution
$P(\Delta)$ for single chain RNAs
(Fig.\ref{asphericity_shape_RNA}-(a)) has a broad peak around
$\Delta\approx 0.3$. This shows that the native-state conformations of
single chain RNA molecules deviate greatly from a sphere.  This
finding is in stark contrast to $P(\Delta)$ in single-chain protein
structures where the peak of the distribution is at $\Delta < 0.1$.\cite{DimaJPCB04} In addition, only $\sim$15\% of single-chain RNA
structures have $\Delta < 0.2$, while in proteins the corresponding
number is $\sim$80\%. This analysis shows that even if native
structures of RNAs are compact ($R_G=5.5N^{1/3}$\AA) they are highly
aspherical.

Because many RNAs are organized as oligomers, we also obtained the
values of $\Delta$ for such structures.  The distribution of $\Delta$
for oligomeric RNAs is also very broad
(Fig.\ref{asphericity_shape_RNA}-(a) middle panel).  Approximately
34\% of the 518 oligomeric RNAs have $\Delta < 0.2$ which shows that
oligomerization in RNA increases the sphericity of the molecule.  This
conclusion is substantiated by analyzing the $R_{\Delta}$
\cite{DimaJPCB04} which is the ratio between the degree of asphericity
of the oligomer and the average ashpericity of the individual chains.
If $R_{\Delta} = 1$ then the oligmers and the chains have the same
asphericity while $R_{\Delta} < 1$ indicates that the oligomer is more
spherical than its components.  Nearly $\sim$60\% of oligomeric RNAs
have $R_{\Delta} \leq 1$.
  
The distribution of the shape parameter, $S$, in single-chain RNAs 
(Fig.\ref{asphericity_shape_RNA}-(b) top panel) 
shows that RNA is mostly prolate because most of the chains have $S > 0$. 
This tendency towards prolate shapes is stronger than in proteins
where $\sim$50\% of single-chains are spherical or nearly so.\cite{DimaJPCB04} 
On the other hand, the complexes of RNA chains found in the PDB structures 
exhibit a bias towards spherical
structures as shown in the peak around $S = 0$ in 
Fig.\ref{asphericity_shape_RNA}-(b) bottom panel. It should be emphasized that there is no systematic dependence of
$\Delta$ or $S$ on N.  A plot of $\Delta$ and $S$ on N shows no correlation whatsoever. The observed variations is directly attributable to
sequence and hence the topology of the folded structure.\\

{\bf Distribution function of radius of gyration can be described by
WLC model: } For the database of RNA molecules, we calculated the
distance distribution, $P(r)$, using the coordinates of the heavy
atoms.  The $P(r)$ functions (Fig.\ref{universality}(a)) for a few RNA
molecules, resemble those obtained using SAXS experiments for compact
RNA molecules.  The value of the persistence length is obtained by
fitting $P(r)$ to Eq.\ref{eqn:fit} in the range $R_G<r<2.5R_G$.  As
can be seen from Fig.\ref{lp} the value of $l_p$ varies between
(5-25)\AA.

If the WLC model correctly describes the distance distribution
function an important prediction follows from Eq.\ref{eqn:fit},
namely, that by replacing $r$ by the dimensionless variable
$x=rl_p/R_G^2$ all the $P(r)$ curves must coincide for $r/R_G>1$.  In
other words, irrespective of the size, sequence or the nature of
interactions that stabilize the native topology, the tail of $P(r)$
($r>R_G$) should superimpose.  Thus, $P(r)$ \emph{should be a function
of only} $l_pr/R_G^2$.  This important prediction is validated in
Fig.\ref{universality}(b) in which a plot of $P(x)$ with
$x=rl_p/R_G^2$ shows that all the structures follow the same
functional form for $x>0.5$ (see \cite{VallePRL05} for the same
analysis performed on the end-to-end distance distribution of DNA).
From this result we conclude that the distance distribution function
of RNA structures are well described by the WLC model. We do not have
any structural basis for this observation.\\

{\bf Persistence length increases with N: } It is remarkable that
$P(r)$ for folded RNA is well described by the WLC model which
accounts only for the bending penalty of a thin elastic material.  The
structural basis for this important finding is not clear.  By fitting
$P(r)$ to Eq.\ref{eqn:fit} for $r/R_G>1$ we find that $l_p$ for folded
structures increases with $N$.  The finding that $l_p$ grows as
$l_p=1.5N^{\alpha}$ with $\alpha\approx1/3$ can be rationalized using
the arguments given below.  A consequence of the sublinear growth of
$l_p$ with $N$ is that the effective contour length for folded RNA
must also grow sublinearly with $N$, i.e.,
$L_{eff}=3\times\left(\frac{5.5^2}{1.5}\right)N^{1/3}\approx
60N^{1/3}$\AA. In the unfolded state we expect the contour length
$L\propto N$. Interestingly, recent single molecular measurements have
also shown that $l_p$ for microtubules depends on the contour length.\cite{PampaloniPNAS06}

The increase in $l_p$ with $N$ is related to the restriction that the
folded states of biomolecules be conformationally less dynamic than
unfolded states.  It is known from polymer physics that if $l_p$ is
fixed and there are no interactions that stabilize a specific
structure then on large scales ($\gg l_p$) the structure would be
intrinsically flexible.  This would mean that spontaneous global
fluctuations of folded RNA would be highly likely due to increase in
conformational entropy.  The requirement that biomolecules should
adopt a near unique native fold which minimizes entropy in the native
basin of attraction (NBA), implies that $l_p$ itself should grow with
$N$.  In contrast, for unfolded RNA, whose conformational entropy is
greater than the structures in the NBA, we expect that $l_p$ should be
independent of $N$ (see Appendix).

The persistence length $l_p$, which determines the flexibility of RNA,
depends on the concentration, shape, and size of counterions.  The
balance of the effective energetics of interactions (stacking
interactions, hydrogen bonding, hydrophobic interaction, and repulsion
between phosphate groups and tertiary interactions) renormalizes
$l_p$.  Let us assume that the interactions are approximated as
pairwise additive and short-ranged $\Delta G\approx
\sum_{|\vec{r}_i-\vec{r}_j|<R_G}\Delta G_{ij}$.  In the presence of
these interactions the persistence length should scale as the range of
the interactions i.e., $l_p\approx R_G\approx N^{1/3}$. The non-local
interactions, which stabilize the folded RNA structures, grow with N
and hence affect $l_p$.  In the absence of interactions that stabilize
the three dimensional fold $l_p$ is determined only by the intrinsic
property of primary sequence and hence should not depend on $N$ (see
Appendix). 

We further rationalize the dependence of $l_p$ on N by noting that
about 54\% of all nucleotides in folded RNA structures are involved in
base pairing (see Fig. 10 in \cite{DimaJMB05}). One possible way,
independent of N, of achieving the 54\% base pairings is to distribute
them over several short duplexes that are stabilized by tertiary
interactions in the native state. Because the tertiary interactions in
RNA are weaker than the base stackings (and other) interactions that
stabilize hairpin-like structures, creation of several short duplexes
is not favorable. Alternatively, it is free energetically more
favorable to create a smaller number of longer stable rigid duplexes that are
stabilized by tertiary interactions to create a nearly spherical
shape.  This strategy seems to operate as N increases as seen in
ribosomes. As a consequence of the presence of large number of rigid duplexes, which reflects
the hierarchical nature of RNA assembly,
$l_p$ increases with N.    In other words, in RNA there is clear separation in energy scales stabilizing secondary and tertiary interactions. Such a hierarchy implies that stiffness itself must be dependent on N.  Because such clear separation in structural organization does not exist in proteins we expect that $l_p$ in proteins must weaker dependence on N (Fig. \ref{lp}). A similar reasoning has been give to explain the growth of $l_p$ with $N$ for
microtubules.\cite{PampaloniPNAS06}\\

{\bf DISCUSSION}\\

{\bf Differences in shapes and packing between proteins and RNA: } It
is difficult to compare, in absolute terms, packing in proteins and
RNA because the nature of interactions that stabilize their native
structures are distinct.\cite{HyeonBC05}  Nevertheless, the Flory
scaling ($R_G\sim aN^{1/3}$) observed in RNA and proteins shows that
both are maximally compact.  For a given $N$, the approximate volume
of RNA is larger than proteins.  The ratio,
$V_{RNA}/V_{PROT}\approx(a_{RNA}/a_{PROT})^3\approx 5.6$ for a fixed
$N$.  This suggests that, in all likelihood, RNA is more loosely
packed than proteins $-$ a conclusion that is in apparent
contradiction with a recent structural analysis.\cite{GersteinJMB05}
Voss and Gerstein based their conclusion on Voronoi construction to
decipher volumes of RNA and specific volume calculations.  They
concluded that ``based on well packed atoms'' RNA is more tightly
packed than proteins.\cite{GersteinJMB05}  The inherent arbitrariness
in assigning volumes to atoms based on Voronoi tessellation of space
and the use of mass in the definition of specific volume obscures
packing effects which should be based on sizes of nucleotides alone.
The present computations show that, based on volume fraction
considerations, RNA is not as compact as proteins as long as $N$ (the
number of nucleotides or the number of aminoacids) is fixed.

The observed differences between shapes of RNA and proteins are
primarily due to the nature of interactions that stabilize the folded
structures of RNA and proteins.  Tertiary structure formation in RNA
must be preceded by substantial neutralization of the negative charges
of phosphate groups.  Condensation of counterions that are
non-specifically bound results in the residual charge on the phosphate
group being less than $\sim -0.1e$ where $e$ is the charge of the
electron.  However, packing in the resulting tertiary fold is
determined not only by interactions involving nucleotides but also by
correlations between counterions. \cite{KoculiJMB04}  The condensation
of a large number of counterions needed to neutralize the charges on
the phosphate groups results in spatial correlation between them.  If
the volume excluded by the counterions is large (for example the
volume of cobalt hexamine is greater than that of Mg$^{2+}$) then
binding of one counterion prevents another one being spatially
adjacent.  These counterion-mediated interactions and their correlation
also inherently affect packing in RNA.  In contrast, packing in the
core of proteins is predominantly determined by interactions between
hydrophobic side chains and their contacts with the protein backbone.
Because of the absence of additional ligands, except in certain cases
like heme proteins, dense packing in proteins is easier to achieve. \\

{\bf Shape fluctuations of proteins and RNA in the ribosome:} The
analysis of shape and flexibility of isolated proteins and RNA gives
insight into packing in isolated biomolecules.  However, in a vast
majority of cases, function requires interactions between two or more
components.  A prime example is the ribosome, a ribonucleoprotein
complex, that plays a central role in protein synthesis.\cite{RamakrishnanNature00_I,SteitzSCI00,YonathCell01,NollerSCI01}
Complexes of both small and large subunits with various antibiotics
have revealed the mechanism of the ribosomal machinery for tRNA
recognition and protein synthesis.
\cite{RamakrishnanCell00,RamakrishnanNature00_II,YonathNature01,SteitzMC02}
The remarkable three dimensional map of entire ribosome (70S)
including three tRNAs and mRNA that shows a snapshot of the
translation process, has also been resolved by cryo-EM techniques at
5.5\AA\ resolution. \cite{NollerSCI01} The binding interface between
30S and 50S subunits, tRNA recognition site in 30S subunit, and
peptidyl transferase site on 50S subunit are all devoid of the
ribosomal proteins.  The cavity is formed at the interface between two
subunits where three tRNA and a string of mRNA can be accommodated.
The structures of $\sim 50$ ribosomal proteins have also been
investigated, giving further insights into the interaction and the
assembly process of the ribosome.\cite{RamakrishnanJMB02,SteitzJMB04}

Comparison of the shapes of the structures in isolation and in the
complex allows us to infer if there are large scale shape changes upon
complexation.  To this end, we analyzed the individual components of
the ribosome as well as each structural domain by using the parameters
that quantify molecular sizes, shapes, and flexibilities of the
individual components.  We used the atomic coordinates from 1GIX (30S
subunit composed of 16S rRNA, 3 tRNA, 1 mRNA, and 20 r-proteins) and
1GIY (50S subunit composed of 23S rRNA, 5S rRNA and 22 r-proteins)
that form an entire ribosome complex upon combination.\cite{NollerSCI01}  The parameters characterizing the structural
components of ribosome are summarized in Table.\ref{table:ribosome}.

{\it r-RNAs :} Each ribosomal RNA (16S, 23S rRNA) can be further
decomposed into several structural domains whose folding is autonomous
even in the absence of ribosomal proteins.\cite{GarrettJMB87,GarrettJMB88,SteitzARB03,WoodsonJMB05}  The
structural features of individual domains of rRNAs in
Fig.\ref{16S_sec}, \ref{23S_sec} are quantified in terms of $R_G$,
$\Delta$, and $S$, with corresponding regions differently colored in
the secondary structure map.  Comparison of $\Delta$ and $S$ values of
rRNA domains (Table.\ref{table:ribosome}) with $P(\Delta)$ and $P(S)$
in Fig.\ref{asphericity_shape_RNA} shows that, except for the 3'm
domain of 16S rRNA, the overall shapes of rRNA domains are
nearly-spherical and slightly prolate ($0<S<0.25$).  Thus, no
significant difference between the overall shape is found in rRNAs
domain in comparison to typical RNA molecules.  However, the
deviations of $R_G$ from the scaling law (Eq.\ref{eqn:Rg_vs_N}),
especially for the domains of 23S rRNA, II, IV, V, VI, show that they
are more extended in size than normal RNA
(Fig.\ref{rRNA_component_Rg}).  We find that the size of the domains
in the 16S rRNA, 5', C, 3'M obeys the scaling law
(Eq.\ref{eqn:Rg_vs_N}).

Because the shape of the fold from each domain is identical to the one
assembled in the intact ribosome, the assembly from extended domains
must occur by a jigsaw puzzle type matching.  The head part of the 16S
rRNA, which is crucial for A, P, E, tRNA binding sites is entirely
composed of the 3'M domain.  The 5' and C domains comprise the body
and the platform part, respectively (see \cite{NollerSCI01} for
terminology).  3'm domain lies at the interface and interacts with
IV-domain of the 23S rRNA when the two subunits dock.  After the rRNA
domains and r-proteins are assembled to form a functional subunit, 50S
subunit is highly spherical ($\Delta=0.05$, $S=-0.01$).  \emph{In
contrast, the 30S subunit is aspherical and prolate} ($\Delta=0.21$,
$S=0.14$).  The acquisition of the spherical shape of the entire
ribosome ($\Delta=0.03$, $S=0.01$) must occur after the folding of two
subunits.  Comparison of the shape of 30S, 50S, and 70S particles
suggests that there is very little alteration in their respective
$\Delta$ and $S$ values upon complexation.  This observation suggests
that these domains probably fold prior to assembly.

Despite their large sizes, the 50S and the 70S particles are
considerably more spherical than the majority of RNA molecules.  The
globular nature of the 50S particle and the 70S complex is surprising
given that the typical RNA complexes are aspherical.  This
asphericity, especially for medium-sized RNA, is the result of coaxial
stacking of helices found in the secondary structures.  The stacking
leads to formation of long helices which are expected to be rigid with
large values of $l_p$.  The 30S subunit, which is highly aspherical
and prolate, fits well with this expectation.  Noller has pointed out
that the ribosome is made up of mostly small helices linked by
flexible bulges and loops. \cite{NollerSCI05}  This observation
applies to the 50S subunit (Fig.\ref{23S_sec}).  However, large-sized
coaxial stackings are dominant in the 16S rRNA, but not in the 23S
rRNA. As a result, the 30S subunit is highly aspherical.  The 70S
complex is highly spherical.  The globularity of the 70S arises
because the 30S subunit fits precisely (despite its high $\Delta$ and
$S$ values (see Table.\ref{table:ribosome})) at the interface with the
50S to create a nearly perfect sphere.

{\it r-proteins :} Similar quantitative analysis can be performed on
the ribosomal proteins.  The values of $R_G$ in some r-proteins
deviate from the scaling law and the shape is generally more biased to
the prolate shape than in the non-ribosomal proteins
(Fig.\ref{r-protein-analysis}).  Ribosomal proteins are mostly
distributed on the back of the interface and the periphery of rRNAs
with some of proteins being anchored deep into the crevices of
rRNA. The anchoring is accomplished using the long tail of peptide
chain composed of positively charged amino acids (ARG, LYS, HIS).\cite{RamakrishnanJMB02,SteitzJMB04}  The unusual topology of
r-proteins prompted us to investigate whether or not the r-proteins
maintain their shape in isolation. We compared the structure of 16
r-proteins complexed in the ribosome ribosome with the isolated
r-protein structures independently determined by X-ray or NMR
available in PDB.  The structural deviation between the isolated and
ribosome-complexed r-proteins is quantified using root mean square
deviation (RMSD).  The structured domains, like $\alpha$-helix and
$\beta$-sheet, are well matched in the isolated protein and in the
complex, but the structural deviation is large in the loop and the
tail regions of the structure.  The structure comparison suggests that
the ordered part of the r-protein is at least well conserved in both
situations.  The disordered tail part is stabilized upon complex
formation inside the crevices of rRNA.\cite{RamakrishnanJMB02,SteitzJMB04} \\

{\bf CONCLUSIONS}\\

In this paper we have shown, by analyzing the available RNA
structures, that $R_G$ can be accurately computed using the celebrated
Flory law.  In contrast to proteins, RNA molecules are considerably
more aspherical with the overall shape being prolate.  The prolate
nature of RNA shapes suggests that their diffusion is intrinsically
anisotropic.  For a given value of N (the number of nucleotides or
amino acids) the persistence length of RNA is considerably larger than
proteins.  These findings suggest that typically RNA is not nearly as
densely packed as proteins even though both are compact in the folded
states.

The structural basis for the success of WLC model in quantitatively
fitting the distance distribution curves for proteins and RNA is not
clear.  It has been appreciated for a long time that elasticity-based
models are appropriate for ds-DNA in monovalent counterions.  The
present findings that $P(r)$ (for $r/R_G>1$) \emph{for compact RNA and
proteins} can be described using polymer models that accounts only for
bending energies is surprising. Our work shows that $l_p$, which is
needed to describe interaction between biomolecules, can be accurately
obtained using the experimentally measurable $P(r)$.  The fit of $P(r)$ to WLC also
shows that $l_p$ increases with $N$.  Such an unusual behavior is, perhaps, related to the need to
minimize entropic fluctuations in the native state.  Suppression of conformational fluctuations in long RNA can achieved by having a small
number of long rigid helices that are stabilized by weak tertiary interactions.  Despite the success of the polymer-based analysis of RNA
structures of varying complexity the microscopic
basis for characterizing for folded
biomolecules using WLC model remains to be established. \\

{\bf APPENDIX}\\

The observation that the persistence length of RNA in the compact
folded states increases as $l_p\approx a_1N^{0.3}$ with $a_1\approx
1.5$\AA\ was rationalized in terms of the restricted conformational
fluctuations in the native state.  A corollary of this interpretation
is that $l_p$ should become independent of $N$ (or the sequence) if
RNA is in the unfolded state.  In this appendix, we adopt an
oversimplified model for the unfolded state of RNA to explicitly show that at
large ($N>40$) $l_p$ indeed does not depend on $N$.

The absence of persistent tertiary structure allows us to describe the
polynucleotide chain as a worm-like chain model.  Such a
coarse-grained description may be an approximate representation of a
single stranded chain made up of one nucleotide (for example polyA).
To verify how $l_p$ changes as $N$ increases we have performed
simulations using WLC which takes into account only the excluded
volume interactions between the beads representing the
nucleotides. The energy function is
\begin{equation}
H=\sum_{i=1}^{N-1}\frac{k_b}{2}(r_{i,i+1}-a)^2+\sum_{i=1}^{N-2}k_a(1-\hat{r}_{i,
i+1}\cdot\hat{r}_{i+1,i+2})+\sum_{i=1}^{N-2}\sum_{j=i+2}^N\frac{k_e}{2}(r_{i.j}-
a)^2\Theta(a-r_{i,j})
\label{eqn:Hamiltonian}
\end{equation}
where $r_{i,j}$, $\hat{r}_{i,j}$ are distance and unit vector between
$i$ and $j$ beads, respectively.  The first term restricts the
extension (or reduction) of bond length around $a$ with
$k_b=2000\epsilon/a^2$ where $\epsilon$ is the unit of energy. The
second term is the bond angle potential that prohibits significant
deviation from the equilibrium value.  We assign $k_a=10\epsilon$.
The last term with $k_e=2000\epsilon/a^2$ takes into account volume
exclusion interaction.  By construction, the homopolymer WLC cannot
form any preferred low energy compact structures.

For this model, whose energy function is given by
Eq.\ref{eqn:Hamiltonian}, we obtained the end-to-end distance
($R_E$) distribution function using Monte Carlo simulations.  Using the  energy function  in Eq.\ref{eqn:Hamiltonian}, 
we generated a large number of equilibrium conformations of the WLC model by 
employing the  pivot algorithm. \cite{BishopJCP91} 
Unlike a standard Monte Carlo methods that generates polymer conformations by moving each monomer 
the pivot algorithm produces a global change in the configuration by 
pivoting the chain around the randomly selected monomer position at
each iteration. The algorithm  enhances  the sampling rate of the available conformational space.  
The acceptance is judged by Metropolis criterion.  

From the ensemble of conformations generated using the pivot algorithm 
we obtained the end-to-end distribution function
$P(R_E)$. The simulated distribution function
$P(R_E)$ can be fit using Eq.\ref{eqn:WLC} from which we obtain $l_p$.
The dependence of $l_p$ on $N$ for the WLC, without the possibility of
forming ordered structures, shows (Fig.\ref{lp_denatured}) that $l_p$
becomes independent of $N$ when $N>40$.  The rise in $l_p$ for $N<40$
is due to the domination of the bending energy (second term in
Eq.\ref{eqn:Hamiltonian}).  For larger values of $N$ the entropic
contributions can compensate for the bending energy and $l_p$
saturates to its intrinsic value. Thus, for WLC with excluded volume
interactions the bending penalty dominates at small $N$ values and the
chain is intrinsically flexible when $N$ is very large.  This
situation is in stark contrast with folded RNA (or proteins) where
$l_p$ grows with $N$.  The increase of $l_p$ as $N$ increases, which
is due to interactions that stabilize RNA, is required to suppress
conformational fluctuations when biomolecules reach the functionally
competent state.  Similar findings are well known for polypeptides
such as polyPro, polyGly, etc. \cite{SchimmelBook}\\

{\bf ACKNOWLEDGMENT}\\

This work was supported in part by a grant from the National Science Foundation 
through NSF CHE-05-14056. 
\\

%\newpage 
\begin{center}
\begin{table}
\fontsize{9}{10pt}\selectfont
\caption{Structural features of the ribosome.}
        \begin{tabular}{|c||c| c c c c c c ||c|c c c c c c |}
        \hline
\multicolumn{1}{|c||}{}&\multicolumn{1}{|c|}{}&\multicolumn{1}{c}{$N$\footnote{$N$ is the number of nucleotides or aminoacids.}}&\multicolumn{1}{c}{$R_G$[\AA]\footnote{The radius of gyration $R_G$ is calculated using Eq.\ref{eqn:RG}.}}&\multicolumn{1}{c}{$\Delta$\footnote{The shape parameters $\Delta$ and $S$ are computed using Eq.\ref{eqn:asphericity}, \ref{eqn:shape}.}}&\multicolumn{1}{c}{$S$}&\multicolumn{1}{c}{$l_p$[\AA]\footnote{$l_p$ is the persistence length.}}&\multicolumn{1}{c||}{RMSD\footnote{The root mean square deviation is the extent of structural deviation of the ribosomal proteins in the complex and in isolation.}[\AA]}&\multicolumn{1}{c|}{}&\multicolumn{1}{c}{$N$}&\multicolumn{1}{c}{$R_G$[\AA]}&\multicolumn{1}{c}{$\Delta$}&\multicolumn{1}{c}{$S$}&\multicolumn{1}{c}{$l_p$[\AA]}&\multicolumn{1}{c|}{RMSD[\AA]}\\
 \hline %\multicolumn{15}{r}{{Continued on next page}} \\
  &{\bf 70S}& 9662& 86.2& 0.03& 0.01&27.1&-&&&&&&&\\
  \cline{2-15}
  &{\bf 30S}& 3915& 66.3& 0.21& 0.14&23.1&-&{\bf 50S}& 5747& 74.3& 0.05& 
-0.01&23.8&-\\
  \cline{2-15}
  &16S & 1519& 65.1& 0.28& 0.21&22.3&-&23S& 2889& 66.4& 0.02& -0.01&23.6&-\\
  \cline{2-15}
   &5'   & 542 & 42.8&0.21& 0.14 &14.9&-&I  &  557& 46.1&0.23 &0.20  &16.5&-\\
   &C    & 352  & 39.8&0.28& 0.25 &14.4&-&II &  736& 56.7&0.08 &0.00  &18.0&-\\
   {\bf r-RNA}&3'M& 484&39.3&0.07& 0.01 &13.4&-&III&  378& 35.4&0.18 &0.12  
&12.3&-\\
   &3'm  & 141 & 45.1&0.66& 1.06 &-\footnote{Persistence length is not reported 
if the correlation coefficient of nonlinear fitting is less than $0.85$.}     
&-&IV &  343& 44.7&0.25 &0.25  &15.6&-\\
   &     &     &     &    &      &      &-&V  &  600& 56.3&0.26 &0.22  &19.4&-\\
   &     &     &     &    &      &      &-&VI &  275& 41.1&0.22 &-0.08 &13.2&-\\
   \cline{2-15}
   &     &     &     &    &      &      & &5S &  123& 32.5& 0.45& 0.59 &10.6&-\\
  \hline
  &S2 & 234  &19.3&0.24 & 0.23 &6.4 &-&L1&224&18.0&0.15&0.09&6.1&5.79\\
  &S3 & 206  &18.3&0.13 & 0.08 &6.1 &-&L2&173&19.1&0.21&0.12&6.0&-\\
  &S4 & 208  &17.6&0.15 & 0.04 &5.9 &-&L3&191&22.7&0.29&0.28&{\bf 
7.1}\footnote{Unusually large values of the parameters ($\Delta$, $S>0.6$, and 
$l_p>7.0$\AA) are given in bold.}&-\\
  &S5 & 150  &16.9&0.28 & 0.29 &5.6 &1.19&L4&189&25.7&{\bf 0.60}&{\bf 0.91}&{\bf 
7.2}&2.90\\
  &S6 & 101  &14.2&0.10 &-0.03 &4.5 &1.17&L5&122&16.9&0.16&0.10&5.7&-\\
  &S7 & 155  &17.5&0.24 & 0.19 &5.6 &0.93&L6&164&19.2&0.39&0.45&5.8&2.06\\
  &S8 & 138  &15.5&0.11 & 0.00 &5.0 &-&L7&128&18.1&0.43&0.53&5.7&0.62\\
  &S9 & 127  &18.2&0.41 & 0.50 &{\bf 21.8}&-&L9&148&25.9&{\bf 0.71}&{\bf 
1.18}&-&2.19\\
  {\bf r-protein}&S10& 98   &18.6&{\bf 0.67}&{\bf 1.08} &{\bf 
23.1}&-&L11&133&16.7&0.30&0.32&5.4&2.00\\
  &S11& 119  &15.0&0.22 & 0.14 &4.7 &-&L12&128&18.1&0.42&0.52&5.6&0.75\\
  &S12& 124  &19.9&0.33 & 0.38 &{\bf 23.7}&-&L13&117&14.2&0.11&0.02&4.6&-\\
  &S13& 125  &22.3&0.46 & 0.56 &{\bf 7.1} &-&L14&122&13.4&0.09&0.02&4.4&-\\
  &S14&  60  &13.8&0.44 & 0.56 &3.9 &-&L15&84&13.8&0.23&0.21&4.2&-\\
  &S15&  88  &13.6&0.19 & 0.08 &4.3 &3.89&L16&138&17.4&0.43&0.56&5.6&17.7\\
  &S16&  83  &12.3&0.13 &-0.06 &3.8 &1.88&L18&113&13.5&0.10&-0.01&4.3&1.84\\
  &S17&  104 &15.5&0.11 & 0.02 &4.9 &4.82&L19&52&10.5&0.09&-0.04&3.0&-\\
  &S18&  73  &12.0&0.08 &-0.04 &3.6 &-&L22&110&16.6&{\bf 0.51}&{\bf 
0.71}&5.3&-\\
  &S19&  80  &12.7&0.08 &-0.02 &3.9 &3.16&L23&76&11.5&0.08&0.00&3.5&4.27\\
  &S20&  99  &17.4&{\bf 0.64}&{\bf 1.02}&5.4 &-&L24&110&15.3&0.08&-0.04&4.9&-\\
  &THX&  24  &7.2 &0.26 & 0.23 &-   &-&L25&89&12.3&0.05&-0.01&3.9&1.41\\
  &&   & &   &    &                & &L29&64&12.8&0.43&0.54&-&-\\
  &&&&&&                           & &L30&60&10.4&0.18&0.09&3.0&-\\
  \hline
  \end{tabular}
 \label{table:ribosome}
\end{table}
\end{center}

\newpage

{\bf FIGURE CAPTIONS}

{\bf Figure} \ref{PDB_RNA}:
Distribution of RNA structures in the Protein Data Bank (PDB) as a function of 
chain length, N.
The arrows show the $N$ values for 16S and 23S ribosomal RNAs, respectively. 
The inset shows the same plot for protein structures. 

{\bf Figure} \ref{RNA_Rg_N_P-P}: (a) Radius of gyration as a function
of $N$.  The straight line is a fit to the data that shows the scaling
law $R_G=5.5N^{0.33}$\AA.  The correlation coefficient if 0.94.  If
data for $N>300$ are neglected we found $R_G=5.6N^{0.33}$ with a
correlation coefficient of 0.92 (fit in green).  Data points inside
the circle, which deviate significantly from the scaling law,
correspond to the structures that are similar to ds-DNA (PDB code:
1H1K). We excluded these structure from the fitting procedure.  For
comparison the plot of $R_G$ as a function of $N$ for 13704 monomeric
proteins are shown in the inset.  The linear line corresponds to
$R_G=3.1N^{0.31}$\AA\ with a correlation coefficient of 0.89.  (b)
Distance distribution of neighboring phosphor atoms along the RNA
backbone.  The distance, $R_{P-P}$ corresponds to separation of the
backbone P atoms between $i^{th}$ and $(i+1)^{th}$ nucleotide where
$i=1,2,\ldots, (N-1)$.

{\bf Figure} \ref{asphericity_shape_RNA}: (a) Distribution of the
asphericity parameter $\Delta$ for RNA.  The top panel corresponds to
single chain, the middle represents single chain in a complex, and the
bottom panel is for the complex.  Large deviation from sphericity is
found in RNA.  (b) Distribution of shape parameters for RNA. The
legend for the three panel is the same as in (a).  RNA molecules in
general are aspherical and prolate like an American football.

{\bf Figure} \ref{universality}: (a) The distance distribution $P(r)$
as a function of $r$ for selected proteins and RNA.  We calculated
$P(r)$ using the coordinates of the folded structures.  The legend at
the bottom gives the PDB codes for which $P(r)$s are shown.  (b)
Dependence of $P(r)$ on the dimensionless variable $x=rl_p/R_G^2$.  If
RNA and proteins can be modeled as WLC then it follows that, for
$x>1$, $P(x)$ should fall on a single line (See Eq.\ref{eqn:fit})
independent of the fold.  The tails of $P(x)$ for $P(r)$ in (a)
practically collapse onto a single curve.  The $\log{P(r)/\beta}$
distributions between dash lines are plotted as a function of
$-\frac{1}{1-x^2}$, which show a nice overlap with the condition,
$\frac{\log{[P(x)/\beta]}}{-1/(1-x^2)}\sim 1$, being satisfied.

{\bf Figure} \ref{lp}: Dependence of $l_p$ on the chain length for
RNA and proteins. The persistence length $l_p$ was computed by fitting
$P(r)$ to Eq.\ref{eqn:fit}.  The lines correspond to
$l_p=1.47N^{0.33}$\AA\ (RNA) and $l_p=1.00N^{0.33}$\AA\
(proteins). There is greater dispersion in the data for proteins than for RNA.
Indeed, the correlation coefficient in the fit for RNA is 0.98 whereas for proteins
it is only 0.79.   Nevertheless, the $l_p$ values for proteins are in the
range inferred from experiments for both peptides and proteinsI. \cite{HofrichterBiophysJ06,HaranPNAS06}

{\bf Figure} \ref{rRNA}: (a) Structural domains of the 16S rRNA. The
corresponding secondary structure at the center is in the same color.
View from interface (left) and back (right) of 16S rRNA assembled by
these structural domains (b) Structural domains of the 23S rRNA. The
organization of the figure is identical to that of 16S rRNA in (a).
The coaxial stackings, are specified as dark lines on the secondary
structures. Molecular graphics images were produced using XRNA and
UCSF Chimera package. \cite{Chimera}

{\bf Figure} \ref{r-protein-analysis}: (a) Radii of gyration ($R_G$)
of the structural domains in 16S (filled circle) and 23S (empty
diamond) rRNAs are plotted as a function of $N$.  Red line
representing $R_G=5.5N^{0.33}$ is drawn to show the deviation of rRNA
domain from the statistics found in usual RNAs.  (b) Plot of $R_G$
against $N$ for ribosomal proteins. Red line represents
$R_G=3.1N^{1/3}$ scaling law found in ``normal'' globular proteins.
Ribosomal proteins (L3, L4, L9, S10, S12, S13, S20) that show a large
deviation from the scaling law are explicitly indicated.  When the
tail part of these proteins are removed, $R_G$ for the r-proteins obey
the Flory scaling law (see open red circles).

{\bf Figure} \ref{lp_denatured}: Persistence length $l_p$ as a
function of $N$ for a WLC model described in the Appendix. This model
may represent a homopolymeric nucleotide at low salt concentrations.
The value of $l_p$ is obtained by fitting the end-to-end distribution
functions $P(R_E)$ that were generated by Monte Carlo simulations (see
Appendix).  An example of $P(R_E)$ as a function of $R_E/L$ for $N=30$
is shown in the inset.  The dependence of $l_p$ in $N$ shows that, for
large $N$, $l_p$ is a constant for a homopolymer chain at low ionic
concentration.

\newpage
\begin{figure}[ht]
\includegraphics[width=5.00in]{PDB_RNA.eps}
\caption{\label{PDB_RNA}}
\end{figure}
\newpage
\begin{figure}[ht]
  \subfigure[]{
    \label{RNA_Rg_N}
    \includegraphics[width=4.15in]{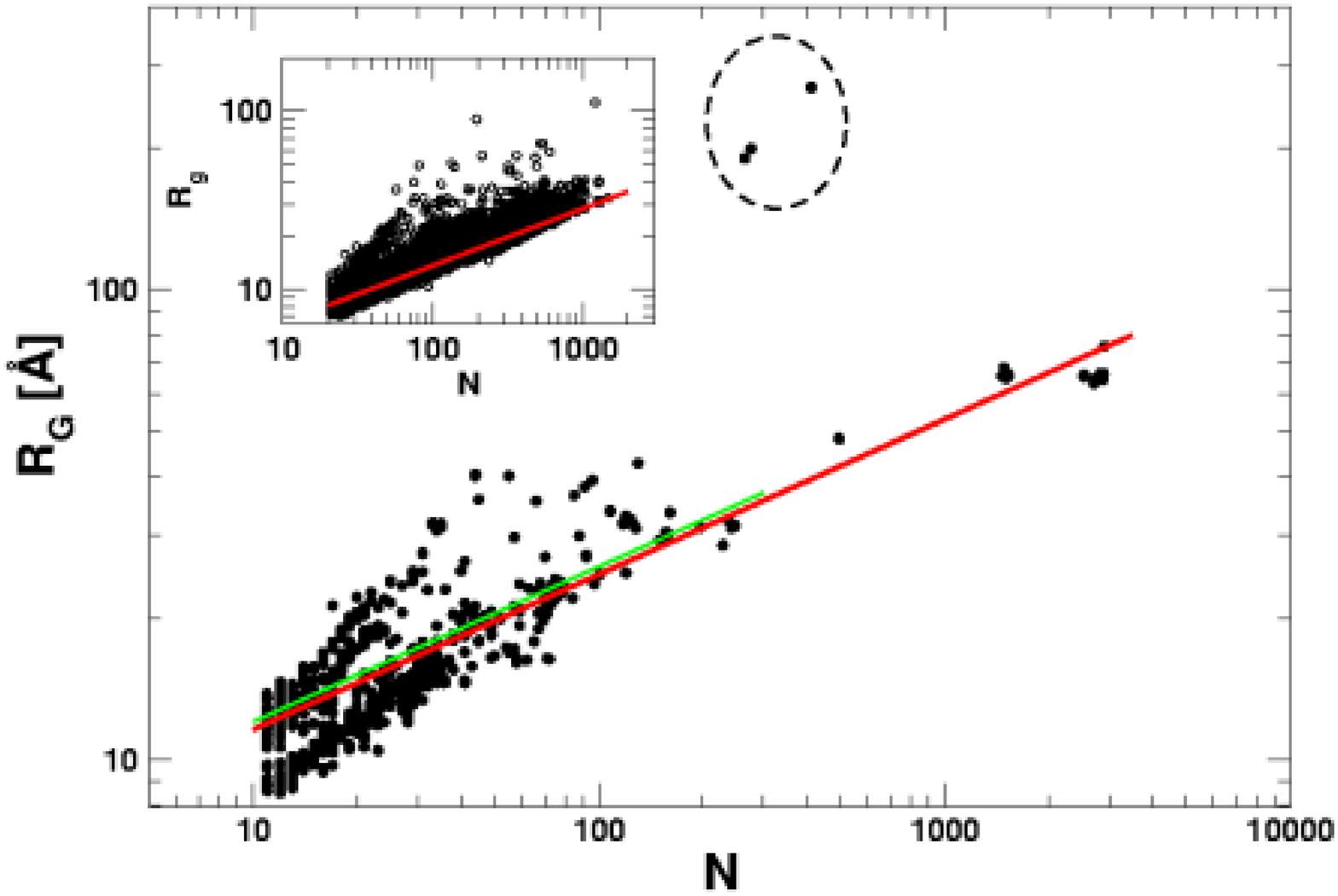}}
  \vspace{0.3in}
  \hfill
  \subfigure[]{
    \label{P-P}
    \includegraphics[width=4.00in]{PP_distribution.eps}}
  \caption{\label{RNA_Rg_N_P-P}}
\end{figure}

\newpage
\begin{figure}[ht]
  \includegraphics[width=6.00in]{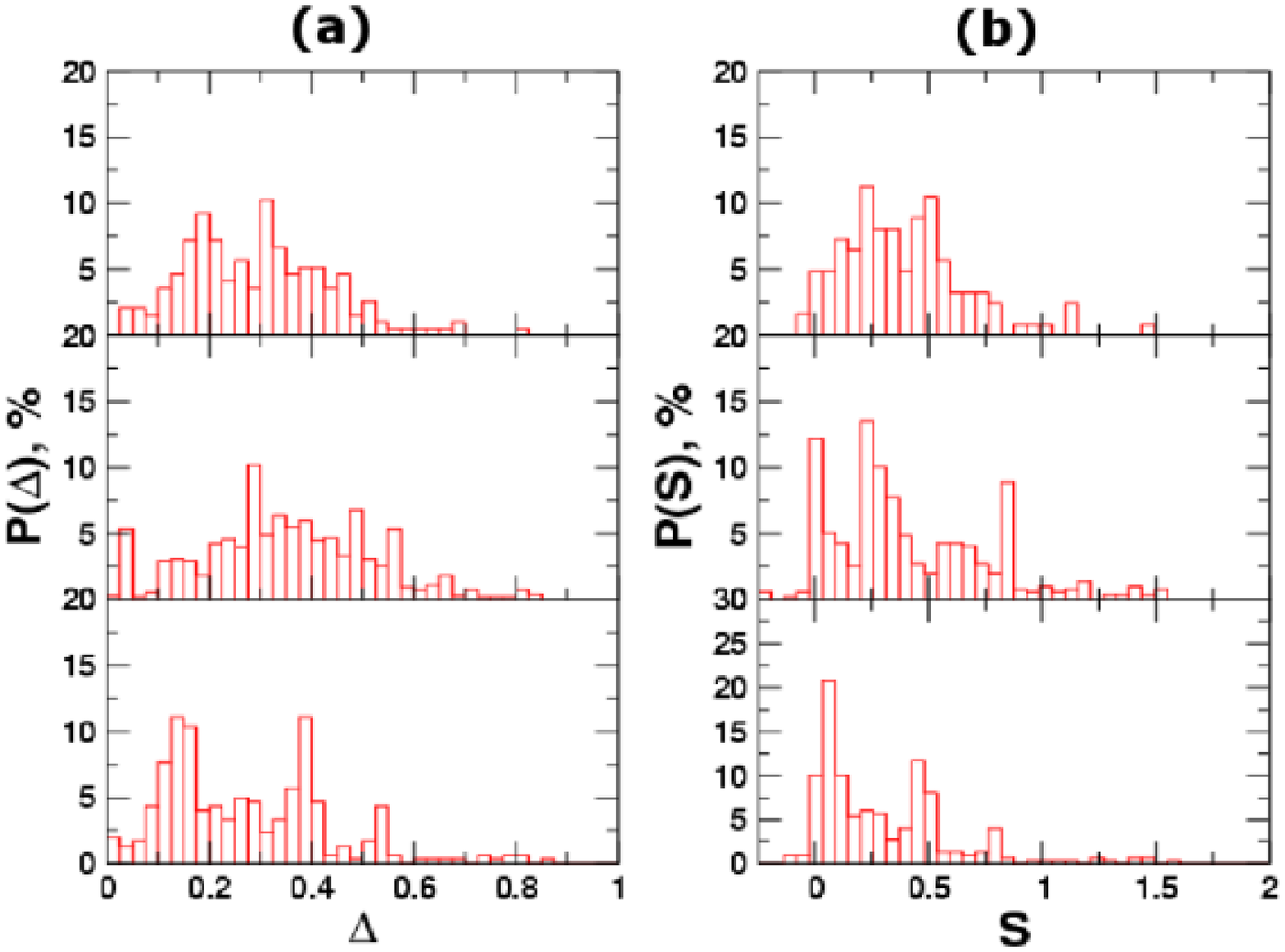}
  \caption{\label{asphericity_shape_RNA}}
\end{figure}

\newpage
\begin{figure}[ht]
  \includegraphics[width=7.00in]{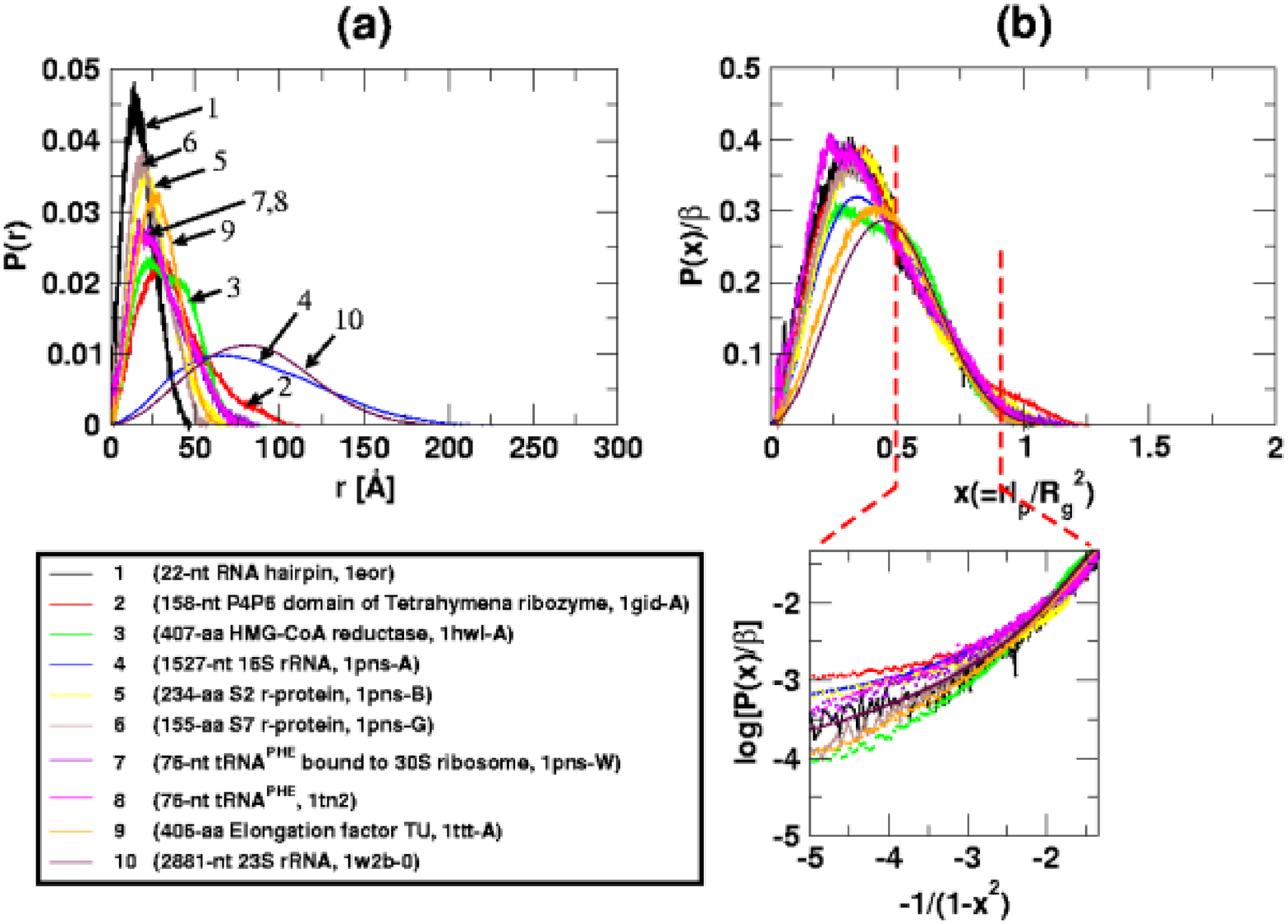}
  \caption{\label{universality}}
\end{figure}

\newpage
\begin{figure}[ht]
\includegraphics[width=3.50in]{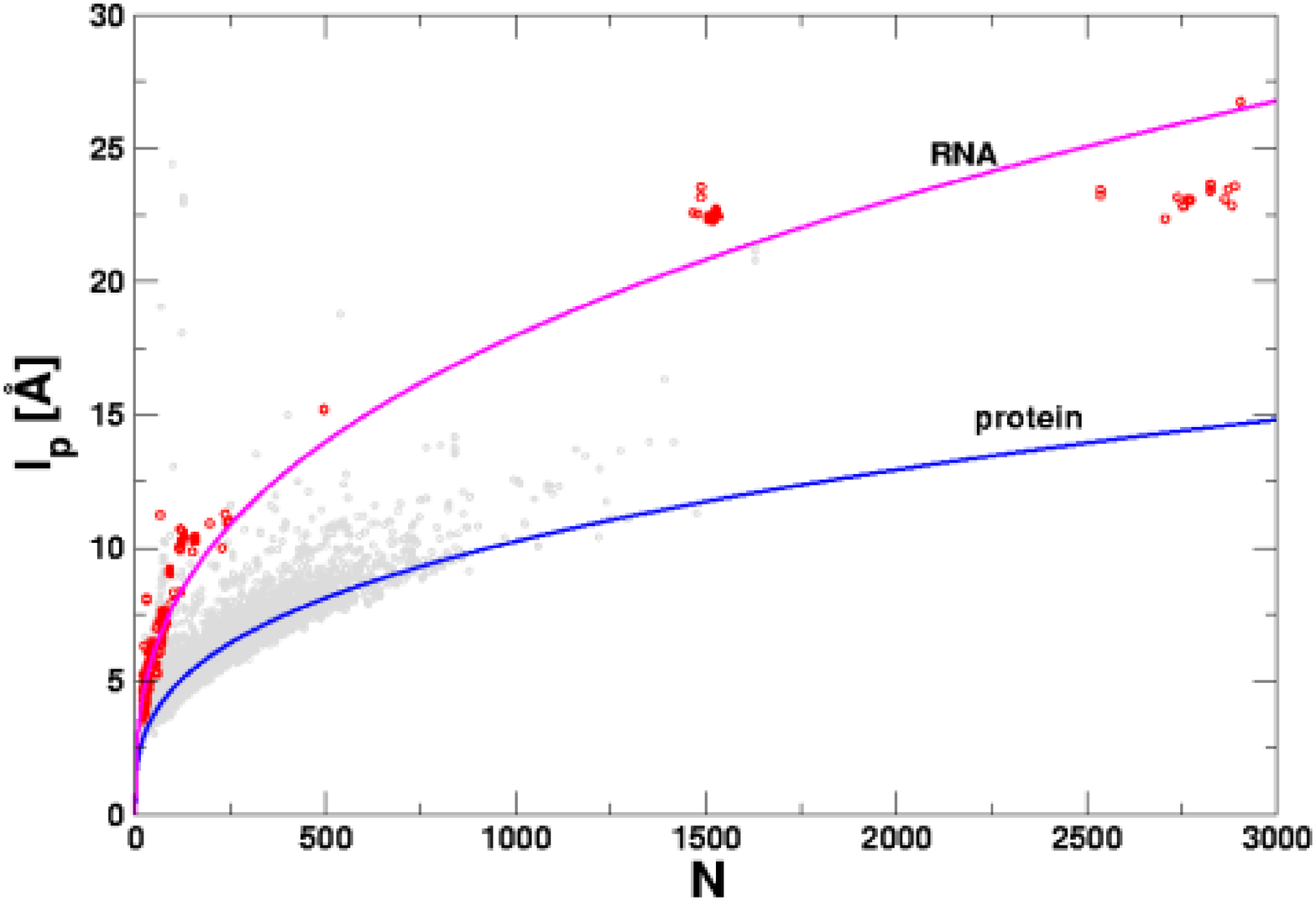}
\caption{\label{lp}}
\end{figure}

\newpage
\begin{figure}[ht]
\centering
  \subfigure[]{
    \label{16S_sec}
    \includegraphics[width=6.00in]{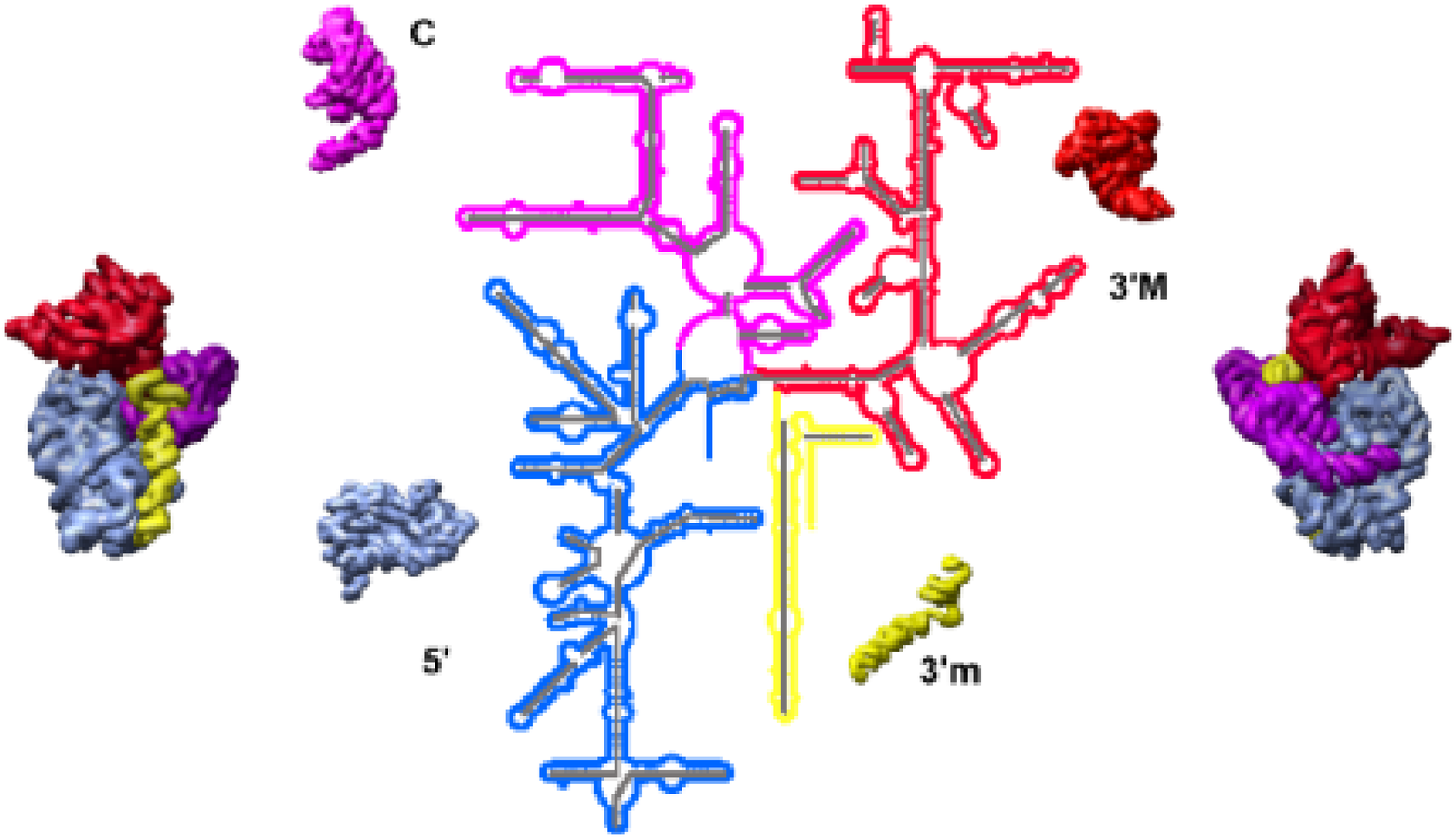}}
    \hfill
  \subfigure[]{
    \label{23S_sec}
    \includegraphics[width=7.00in]{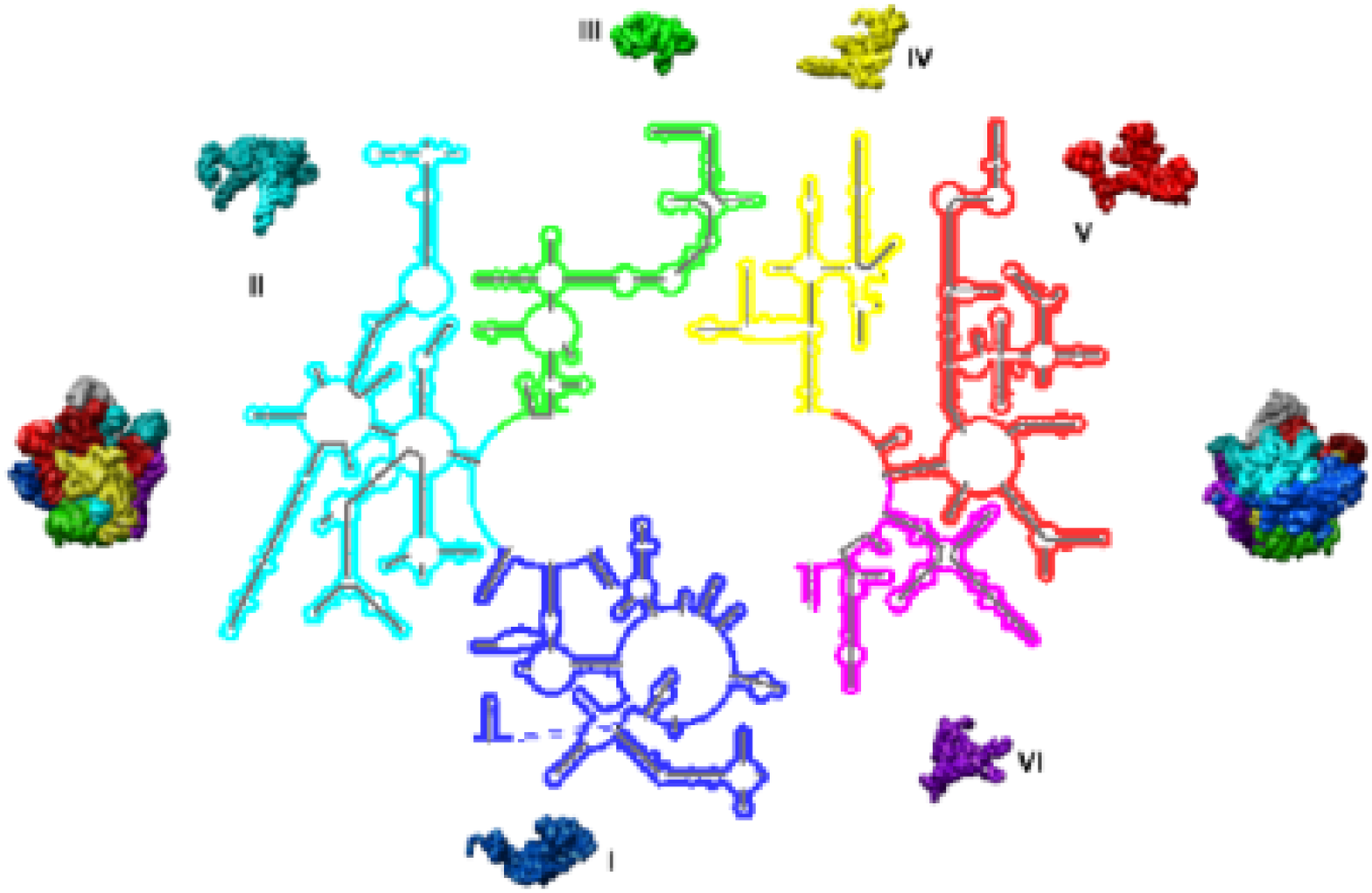}}
    \hfill
\caption{\label{rRNA}}
\end{figure}
\newpage
\begin{figure}
  \subfigure[]{
    \label{rRNA_component_Rg}
    \includegraphics[width=3.00in]{rRNA_component_Rg.eps}}
  \hfill
  \subfigure[]{
    \label{1gix1giy_Rg}
    \includegraphics[width=3.00in]{1gix1giy_Rg.eps}}
  \caption{\label{r-protein-analysis}}
\end{figure}

\newpage
\begin{figure}[ht]
\includegraphics[width=4.00in]{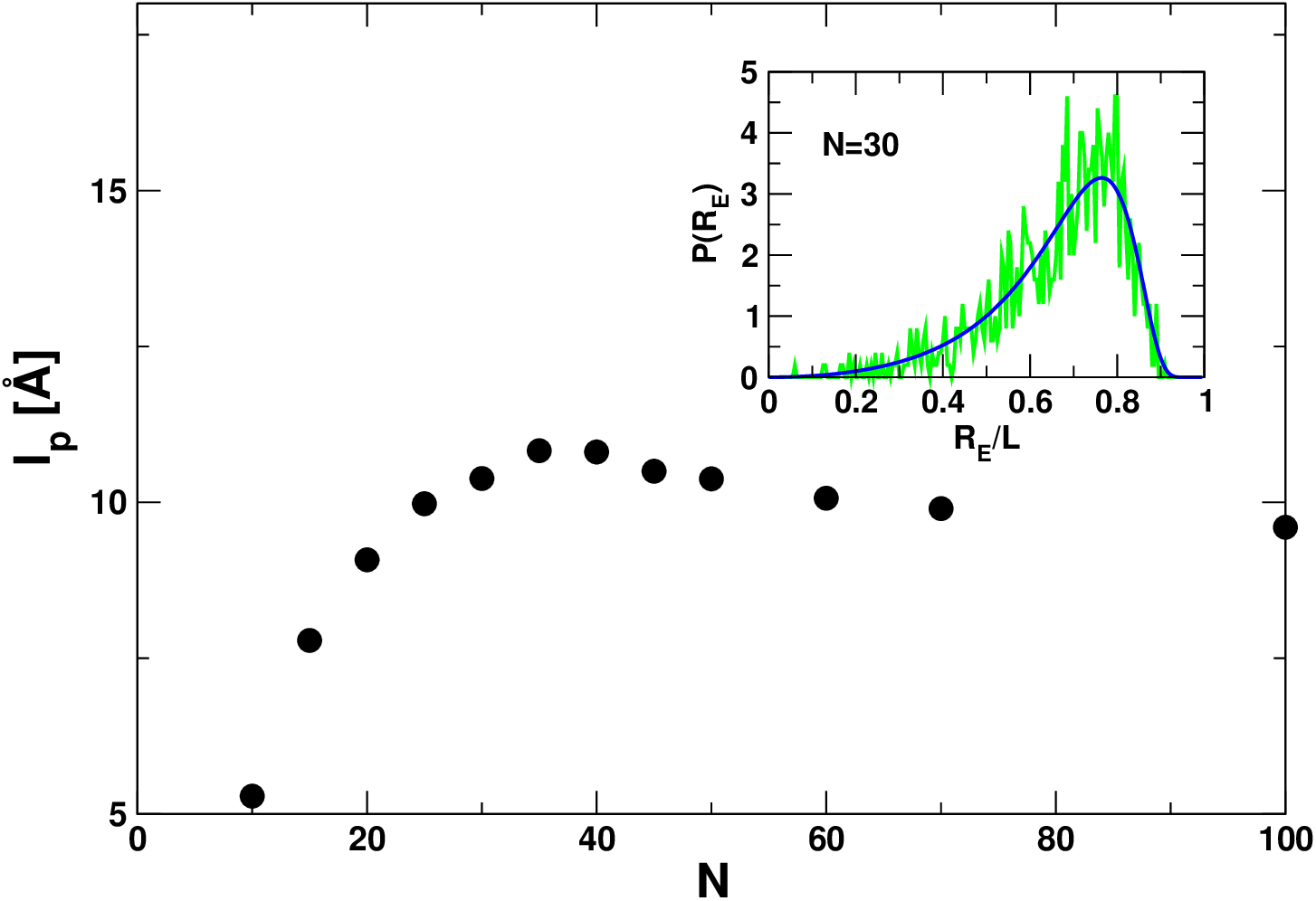}
\caption{\label{lp_denatured}}
\end{figure}

\end{document}